\documentclass[twocolumn,english,aps,prb,superscriptaddress,natbib,bibnotes,amsmath,amssymb,floatfix,groupedaddress,footinbib]{revtex4-2}

\usepackage[colorlinks=true,citecolor=blue,linkcolor=magenta]{hyperref}

\usepackage[markup=nocolor, authormarkupposition=left]{changes} 

\usepackage{soul}
\usepackage[utf8]{inputenc}
\usepackage[english]{babel}
\usepackage{amsmath,amsfonts,amssymb}
\usepackage[T1]{fontenc}
\usepackage{url}

\usepackage{amsmath}
\usepackage{siunitx}
\usepackage[version=4]{mhchem}
\usepackage{amsfonts}
\usepackage{amssymb}

\usepackage{epstopdf}
\usepackage{graphicx}
\graphicspath{{./Figures/}}

\usepackage{changes}

\newcommand{\SiN}[0]{Si$_3$N$_4$}
 \newcommand{\chitwo}{$\chi^{(2)}$}

\begin{document}

\title{Self-organized spatiotemporal quasi-phase-matching in microresonators}

\author{Ji Zhou$^{1,\ast}$, 
        Jianqi Hu$^{1,\ast,\dagger}$, 
        Marco Clementi$^{1,\ast}$, 
        Ozan Yakar$^1$, 
        Edgars Nitiss$^{1}$, 
        Anton Stroganov$^{2}$ and Camille-Sophie Brès$^{1,\ddag}$}
\affiliation{
$^1$Photonic Systems Laboratory\char`,{} École Polytechnique Fédérale de Lausanne\char`,{} 1015 Lausanne\char`,{} Switzerland.\\
$^2$LIGENTEC SA\char`,{} EPFL Innovation Park\char`,{} 1024 Ecublens\char`,{} Switzerland.
}

\date{\today}

\begin{abstract}
Quasi-phase-matching (QPM) is a widely adopted technique for mitigating stringent momentum conservation in nonlinear optical processes such as second-harmonic generation (SHG). It effectively compensates for the phase velocity mismatch between optical harmonics by introducing a periodic spatial modulation to the nonlinear optical medium. Such a mechanism has been further generalized to the spatiotemporal domain, where a non-stationary spatial QPM can induce a frequency shift of the generated light.
Here we demonstrate how a spatiotemporal QPM grating, consisting in a concurrent spatial and temporal modulation of the nonlinear response, naturally emerges through all-optical poling in silicon nitride microresonators. Mediated by the coherent photogalvanic effect, a traveling space-charge grating is self-organized, affecting momentum and energy conservation, resulting in a quasi-phase-matched and Doppler-shifted second harmonic. Our observation of the photoinduced spatiotemporal QPM expands the scope of phase matching conditions in nonlinear photonics.
\end{abstract}

\maketitle

\section*{Introduction}\label{intro}
\noindent Efficient optical frequency conversion via nonlinear light-matter interaction generally requires both energy and momentum conservation among participating photons. Momentum conservation (i.e., phase velocity matching) is, however, often hindered by the chromatic dispersion of nonlinear optical media \cite{Boyd2020NonlinearOptics4th}.
An effective strategy to overcome this limitation is quasi-phase-matching (QPM), initially proposed by Armstrong and coworkers in 1962 \cite{Armstrong1962InteractionsLightWaves}, which introduces an ordered spatial modulation of the nonlinear susceptibility to compensate the momentum mismatch between optical waves. 
To date, QPM is a widely used technique to enable second-order (\chitwo) nonlinear processes like second-harmonic generation (SHG), sum/difference-frequency generation, and spontaneous parametric down conversion. 
Highly efficient SHG has been achieved by periodic poling of non-centrosymmetric materials such as lithium niobate (LN) and KTiOPO$_4$ (KTP) bulk crystals \cite{Fejer1992Quasiphasematched,Karlsson1997Electricfieldpoling,Hum2007Quasiphasematching}, as well as integrated thin-film LN waveguides and microresonators \cite{Wang2018Ultrahighefficiencywavelength,Lu2019Periodicallypoledthin}. 
In recent years, following the work carried in doped silica fibers \cite{Dianov1991Photoinducedeffectsoptical, Anderson1991Modelsecondharmonic, Krol1991Photoinducedsecondharmonic,Margulis1995Imagingnonlineargrating}, there has also been a growing interest in realizing such \chitwo\ functionalities in integrated centrosymmetric media, such as silicon \cite{Timurdogan2017Electricfieldinduced, Singh2020Broadband200nm}, and silicon nitride (\SiN) \cite{Levy2011Harmonicgenerationsilicon,Porcel2017Photoinducedsecond, Billat2017Largesecondharmonic, Hickstein2019Selforganizednonlinear,  Nitiss2020FormationRulesDynamics, Lu2020Efficientphotoinducedsecond, Nitiss2022Opticallyreconfigurablequasi, Hu2022Photoinducedcascaded, Clementi2023chipscalesecond}. Despite lacking an intrinsic second-order nonlinearity, these materials can be endowed with an effective \chitwo\ by breaking the inversion symmetry through the application of electric or optical fields. 

Amidst these, all-optical poling (AOP) has been an effective means to induce \chitwo\ nonlinearity. Requiring only moderate power, it allows for the optical inscription of \chitwo\ and self-configuration of QPM. This process is enabled by the coherent photogalvanic effect (CPE), wherein the interference among multi-photon absorption processes results in the emergence of coherent currents, and yields the inscription of a static electric field inherently supporting the required QPM condition \cite{Dianov1995Photoinducedgenerationsecond,Yakar2022GeneralizedCoherentPhotogalvanic}. In \SiN\ microresonators, AOP has allowed for SHG with high conversion efficiency (CE) \cite{Lu2020Efficientphotoinducedsecond} and broadband reconfigurability \cite{Nitiss2022Opticallyreconfigurablequasi}.

\begin{figure*}[htbp]
\centering
\includegraphics[width=1\linewidth]{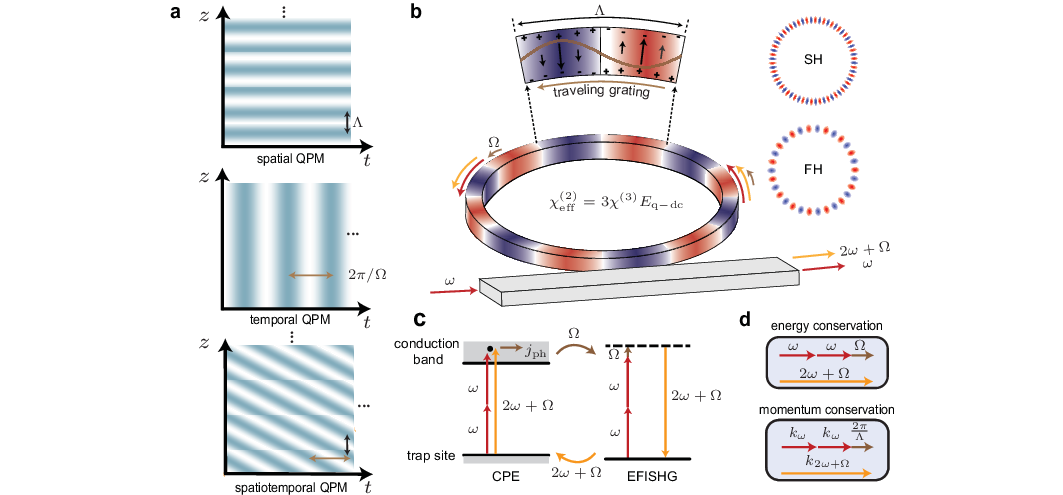}
\caption{
    \textbf{Self-organized spatiotemporal quasi-phase-matching in \SiN \ microresonators}.
    \textbf{a}, Various quasi-phase-matching (QPM) schemes. Top: spatial QPM with a period of $\Lambda$ can compensate for the wavevector mismatch of $2\pi/\Lambda$, e.g., in a standard second-harmonic generation (SHG) process \cite{Karlsson1997Electricfieldpoling,Wang2018Ultrahighefficiencywavelength}. Middle: temporal modulation with a period of $2\pi/\Omega$ can shift the optical frequency by $\Omega$, e.g., in an electro-optic modulation process \cite{Houtz2009WidebandEfficientOptical}. Bottom: Spatiotemporal QPM combining both spatial and temporal modulation can simultaneously mitigate momentum and energy mismatch, as demonstrated in processes such as high-harmonic generation \cite{Bahabad2010Quasiphasematching,Shoulga2023Allopticalspatiotemporal} and photoinduced SHG within this study. 
    \textbf{b}, Spatiotemporal QPM based on a self-organized traveling $\chi^{(2)}$ grating for SHG in a \SiN \  microresonator. When the doubly resonant condition is met, the pump at the fundamental-harmonic (FH) frequency $\omega$ inscribes a $\chi^{(2)}$ grating traveling at the frequency of $\Omega$ and leads to the generation of the second-harmonic (SH) at the frequency of $2\omega+\Omega$.
    The upper zoom-in shows a period of transverse quasi-DC electric field mediated by separated alternating space-charges, inducing a traveling $\chi^{(2)}$ grating by $\chi^{(2)}_{\mathrm{eff}}=3\chi^{(3)}E_{\mathrm{q-dc}}$.
    The right insets show the artistic illustration of optical mode profiles at the FH and SH bands. 
    \textbf{c}, Photoinduced SHG from the interplay between the coherent photogalvanic effect (CPE) and the electric-field-induced SHG (EFISHG) effect. In the CPE, the interference between two pump photons and one SH photon absorption process generates an anisotropic coherent current $j_{\mathrm{ph}}$, which allows for the inscription of a quasi-dc field $E_{\mathrm{q-dc}}$. The $E_{\mathrm{q-dc}}$ enables the quasi-phase-matched EFISHG process for efficient generation of the SH, which in turn enhances the CPE, forming a self-sustaining positive feedback loop. 
    \textbf{d}, Momentum and energy conservation diagrams. The spatial and temporal modulation of the traveling \chitwo \ grating affects the momentum and energy conservation, respectively.}
\label{fig_1}
\end{figure*}

While the spatial properties of photoinduced QPM have been extensively investigated \cite{Hickstein2019Selforganizednonlinear,Nitiss2020FormationRulesDynamics,Nitiss2022Opticallyreconfigurablequasi}, the dynamics of the process have not been explored, despite a recent work bringing forward the hypothesis of non-stationary solutions by considering photoinduced SHG as optical parametric oscillations \cite{Lu2021ConsideringPhotoinducedSecond}. Indeed the nature of the AOP process could entail an additional temporal behavior, similar to spatiotemporal QPM, which was proposed and demonstrated in high-harmonic generation \cite{Zhang2007Quasiphasematching,Bahabad2010Quasiphasematching, Shoulga2023Allopticalspatiotemporal}. In this context, the momentum mismatch compensation through the spatial modulation is complemented by a temporal modulation that analogously compensates for an energy mismatch (Fig. \ref{fig_1}a).

In this work we demonstrate, for the first time, the spatiotemporal dynamics of AOP in \SiN \ microresonators. 
Mediated by the CPE, the photoinduced electric field in turn enhances SHG, also known as the electric-field-induced SHG (EFISHG) effect. Remarkably, we find that the dynamics of resonant AOP yields a temporal modulation of the photoinduced nonlinearity, associated with the spatial one, which altogether can be regarded as a traveling \chitwo\ grating that moves indefinitely alongside the microring circumference. We investigate the spatial properties of such spatiotemporal QPM by two-photon microscopy (TPM) imaging, while its temporal structures is characterized by self-homodyne and self-heterodyne measurements with a frequency-doubled reference. From the latter, we observe an additional frequency (energy) offset from the typical SHG process, which can be interpreted as a Doppler shift from the interaction with a traveling $\chi^{(2)}$ grating. Our findings shed light on the physics of resonant AOP, providing new concrete experimental evidence for the spatiotemporal QPM \cite{Bahabad2010Quasiphasematching} and establishing a comprehensive model for photoinduced nonlinear processes in resonant systems.

\section*{Results}
\noindent \textbf{Principle of self-organized spatiotemporal QPM in microresonators.}
\noindent We begin by describing the photoinduced SHG in a \SiN \ microresonator (Fig.~\ref{fig_1}b). When both the pump and its second-harmonic (SH) are doubly resonant, the initial weak light within the SH resonance seeds the CPE, and the interference between single- and two-photon absorption processes generates an anisotropic coherent current \cite{Anderson1991Modelsecondharmonic, Dianov1991Photoinducedeffectsoptical, Yakar2022GeneralizedCoherentPhotogalvanic}: 
\begin{equation}j_{\mathrm{ph}}=\beta(E_{\mathrm{pump}}^{*})^2E_{\mathrm{SH}}e^{i\Delta kR\phi}e^{-i\psi_{\mathrm{ph}}}+c.c.
	\label{jph}
\end{equation}
where $\beta$ and $\psi_{\mathrm{ph}}$ are the photogalvanic coefficient and the interaction phase, respectively. $E_{\mathrm{pump}}$ and $E_{\mathrm{SH}}$ are the optical fields of the pump and its SH, with $\Delta k=k_{\mathrm{SH}}-2 k_{\text {pump}}$ the wavevector mismatch between them. $\phi$ denotes the azimuthal angle along the circumference of the microresonator with a radius of $R$. Both $^{*}$ and c.c. stand for complex conjugate.

The separation of charges via coherent current leads to the creation of a photoinduced electric field, while the concurrent drift current results in its decay. In the steady state, the photoinduced field can be expressed as $E=-j_{\rm{ph}}/\sigma(I_{\mathrm{pump}}, I_{{\mathrm{SH}}})$, with $\sigma$ the conductivity which depends on the pump and SH light intensity $I_{\mathrm{pump}}$ and $I_{\mathrm{SH}}$, respectively. Notably, the wavevector mismatch between the inscribing optical fields renders a spatial modulation of the photoinduced electric field with a period $\Lambda=2\pi/\Delta k$. 
Meanwhile, the electric field also imparts an effective \chitwo\ nonlinearity, i.e., $\chi^{(2)}_{\mathrm{eff}}=3\chi^{(3)}E$, thus giving rise to EFISHG. The generated SHG further enhances the CPE (Fig.~\ref{fig_1}c), forming a positive feedback loop that leads to the growth of SH, ultimately limited by the increase of the photoconductivity $\sigma$. Remarkably, when the SHG process is interrupted, the displaced charges maintain their spatial distribution owing to the insulating nature of \SiN \ in the absence of excited carriers. The inscription of a long-lasting nonlinear grating is confirmed by electric-field-sensitive etching \cite{Margulis1995Imagingnonlineargrating} and TPM imaging \cite{Nitiss2020FormationRulesDynamics,Nitiss2022Opticallyreconfigurablequasi}.

To gather insights into the AOP dynamics in microresonators, we model the temporal dynamics of doubly resonant SHG with coupled-mode equations (CME) (see Eqs.~(\ref{CME}) in Methods). The theoretical analysis provides two important results for efficient SHG by AOP (see Supplementary Note I):
\begin{itemize}
    \item The inscribed $\chi^{(2)}$ grating is stable only when the generated SH ($\omega_{\text{SH}}$) lays on the blue side of the SH resonance ($\omega_{s}$), i.e., when the detuning of the SH light satisfies:
    \begin{equation}
        \omega_{s}-\omega_{\text{SH}} \approx \delta^{\prime}_{s} <0 \label{deltas}.
    \end{equation}
    where $ \delta_{s}^{\prime}=\omega_{s}- 2\omega_{\text{pump}} $ with $\omega_{\text{pump}}$ the pump frequency. The approximation is valid as $\omega_{\text{SH}}=2\omega_{\text{pump}}+\Omega$ with $\Omega/2\pi$ in the sub-kHz range, according to both the theoretical estimation and experimental observation.
    \item The photoinduced grating exhibits a temporal oscillation with an angular frequency:
    \begin{equation}
        \Omega \approx \frac{\kappa _s}{2 \delta _{s}^{\prime} \tau} \label{Omega},
    \end{equation}
    where $\kappa_s$ is the linewidth of the SH resonance and $\tau$ is the intensity-dependent grating lifetime.
\end{itemize}
These results, together with the spatial periodic modulation, can be interpreted as a traveling $\chi^{(2)}$ grating, as schematically shown in Fig.~\ref{fig_1}b. Such simultaneous spatial and temporal modulation of the \chitwo\ nonlinearity now arises from a \textit{quasi}-static electric field in the microresonator:
\begin{equation}
    E_{\text{q-dc}}\sim e^{i\frac{2\pi}{\Lambda}R\phi-i\Omega t},
    \label{Ai phase}
\end{equation}
where the electric field moves along the ring circumference with a phase velocity $v=\Omega\Lambda/2\pi$. Intuitively, the movement of the grating can be understood as a consequence of its self-organized nature: as the SH field is a source of the photoinduced electric field and vice versa (see Eqs.~(\ref{CME}) in Methods and also Supplementary Note I), these two fields can share an arbitrary phase relation with respect to the pump field, which takes the form of the time-dependent phase shift $\varphi(t)=\Omega t$. In other words, the grating can move due to the existence of an unconstrained degree of freedom between pump, SH field and photoinduced electric field.

An important consequence of these findings is that the traveling grating influences both momentum and energy matching, respectively related to the spatial and temporal modulation, as illustrated in Fig.~\ref{fig_1}d.
While spatial QPM is well understood \cite{Armstrong1962InteractionsLightWaves}, the energy modification related to its temporal counterpart can be regarded as a Doppler shift imparted by the traveling grating to the generated field. More generally, the phenomenology described here can be set in the framework of the generalized spatiotemporal QPM \cite{Berger1998NonlinearPhotonicCrystals,Bahabad2010Quasiphasematching}, where the photoinduced SHG in a microresonator involves the scattering of the pump in a nonlinear photonic crystal. In our spatiotemporal QPM case (Fig.~\ref{fig_1}a bottom), during AOP the CPE organizes a traveling nonlinear grating that effectively doubles the frequency of the pump, with an additional frequency translation of $\Omega$ satisfying energy conservation.

\begin{figure*}[pbt]
	\centering
	\includegraphics[width=1\linewidth]{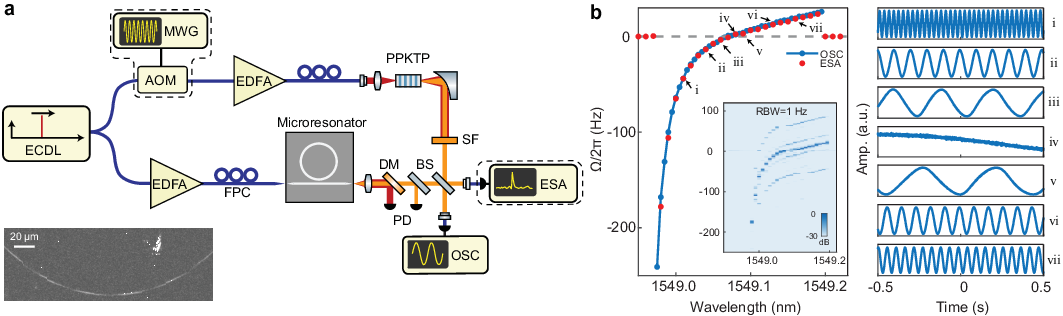}
	\caption{\textbf{Characterization of the photoinduced second-harmonic frequency offset}. 
    \textbf{a}, Experimental setup for measuring the frequency offset between second-harmonics generated in a \SiN \ microresonator and a standard frequency doubler. ECDL: external-cavity diode laser; MWG: microwave generator; AOM: acousto-optic modulator; EDFA: erbium-doped fiber amplifier; PPKTP: periodically poled KTiOPO$_4$ crystal; SF: spectral filter; FPC: fiber polarization controller; DM: dichroic mirror; PD: photodetector; BS: beam splitter; ESA: electrical spectrum analyzer; OSC: oscilloscope.
    Both self-homodyne measurements of temporal oscillations using an OSC (excluding dashed boxes) and self-heterodyne measurements of frequency spectra using an ESA are performed. The inset showcases a \chitwo \ nonlinear grating obtained using the two-photon microscopy after the microresonator is poled near the pump resonance at 1549.0 nm. The participating SH mode is TE$_{30}$ inferred from the measured grating period. 
    \textbf{b}, Left: measured frequency offsets (blue dot-lines: self-homodyne; red dots: self-heterodyne) as functions of the pump wavelength when tuning the laser wavelength across the resonance at 1549.0 nm from blue to red side. The inset shows the spectral map on a logarithmic scale obtained from the self-heterodyne measurement with a 1 Hz resolution bandwidth (RBW) in ESA. 
    Right: temporal oscillation traces measured by the self-homodyne technique for the data points indicated on the left panel (i-vii).}\label{fig_2}
\end{figure*}

\medskip
\noindent\textbf{Temporal characterization of spatiotemporal QPM.}
To verify our theoretical predictions, we experimentally investigate the temporal aspect of self-organized QPM. Fig.~\ref{fig_2}a shows the experimental setup (see Methods) used for measuring the frequency offset between the photoinduced SH generated from a \SiN \ microresonator and a reference SH generated from an external crystal. The \SiN \ microresonator used in this study has a radius of 158 $\mu$m (see Methods), and many of its TE$_{00}$ resonances at the telecommunication C-band were found capable to generate SHG via AOP \cite{Nitiss2022Opticallyreconfigurablequasi}. We perform nonlinear self-heterodyne measurements by recording the optical beating between the two generated SH fields in an electrical spectrum analyzer (ESA) (see Methods). 
This technique offers a 1 Hz resolution (defined by the resolution bandwidth (RBW) of the ESA) in measuring the possible frequency offset, and allows for the unambiguous determination of the exact photoinduced SH frequency. If only purely spatial QPM takes place, the beatnote frequency will be exactly twice of the modulation frequency owing to the SHG process. Otherwise, the presence of the traveling nonlinear grating would impart an additional frequency offset. 

The left panel of Fig.~\ref{fig_2}b presents the steady state frequency offsets experimentally measured at different pump detunings for an AOP instance. In this case, the pump is tuned into the FH resonance near 1549.0 nm from the blue to the red side, and the participating SH mode is TE$_{30}$ mode identified by TPM imaging (Fig.~\ref{fig_2}a inset). The offset frequencies are retrieved from the ESA spectra (inset). At each detuning, a clear single tone in the range of a few hundred Hz is measured for the frequency offset $\Omega/2\pi$, while the electronic pickup signal is much weaker. Notably, the beatnote signal measured when SH is not generated from the microresonator, is resulted from the beating of the SHG of the weak residual pump and the frequency-shifted local oscillator in the crystal path only. This corresponds to the exact SHG process, serving as a frequency calibration for the offset frequency measurements. 
In addition, we repeat the experiment in a self-homodyne scheme and record the temporal traces in an oscilloscope (see Methods). In this case, we observe temporal oscillations after beating the SH light, which precisely replicates the trend recorded with the ESA, as shown in Fig.~\ref{fig_2}b. We note that the frequency offset trace here crosses zero near 1549.08 nm, deviating from the theoretical prediction (Eq.(\ref{Omega})). We attribute this behavior to possibly the perturbation from a neighboring SH mode to the SHG process (see Supplementary Note II).

\begin{figure*}[hbpt]
	\centering
	\includegraphics[width=0.97\linewidth]{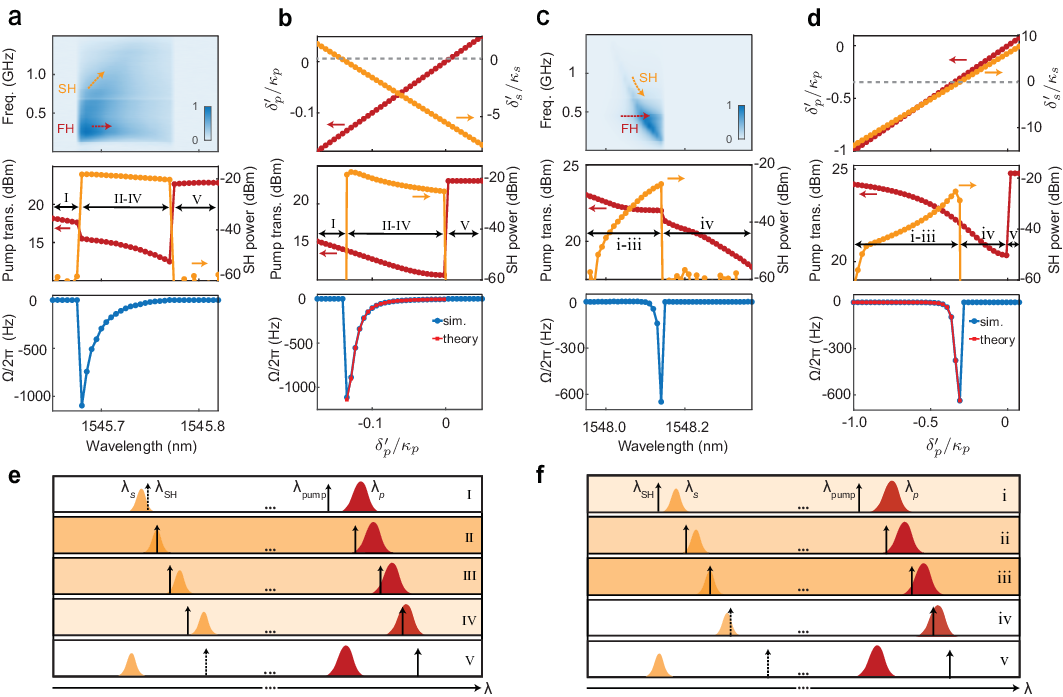}
    \caption{\textbf{Dynamics of photoinduced second-harmonic generation}. 
    \textbf{a},\textbf{c}, Experimental investigations of photoinduced SHG processes for (\textbf{a}) the leading and (\textbf{c}) trailing  cases. They correspond to the scenarios where the generated SH power decreases and increases with the pump tuning into its resonance, respectively. Top: effective detunings of the pump and SH resonances measured by a vector network analyzer; Middle: Measured pump transmission and generated SH power; Bottom: Measured frequency offset as a function of the pump wavelength. \textbf{b},\textbf{d}, Numerical simulations of photoinduced SHG processes for the leading (\textbf{b}) and trailing (\textbf{d}) cases.
    Top: simulated effective detunings of the pump and SH resonances (the grey dashed lines denote $\delta^{\prime}_{s} =0$); Middle: simulated pump transmission and generated SH power; Bottom: simulated and theoretically predicted frequency offsets as functions of the effective detuning. 
    \textbf{e},\textbf{f}, Schematics of dynamic doubly resonant conditions for the leading (\textbf{e}) and trailing (\textbf{f}) cases. 
    During pump wavelength tuning from stage $\rm{I}$ to $\rm{IV}$ or $\rm{i}$ to $\rm{iv}$, the pump ($\lambda_{\text{pump}}$) stays thermally locked to the blue side of its resonance ($\lambda_p$), while the SH resonance ($\lambda_s$) varies its relative position from the generated SH wavelength ($\lambda_{\text{SH}}$). The leading (trailing) case corresponds to the scenario where the red-shift rate of the SH resonance is two times larger (smaller) than that of the FH resonance when tuning the pump into its resonance. The intensity of the background colour indicates the power of the generated SH. At stage $\rm{V}$ and $\rm{v}$, the pump exists the resonance.}\label{fig_3}
\end{figure*}

\medskip
\noindent\textbf{Dynamics of photoinduced SHG.}
To delve deeper into the dynamics of self-organized spatiotemporal QPM in the \SiN\ microresonator, we perform experiments for several other resonances at C-band that support efficient SHG (see Methods). The experimental results can be generally divided into two main cases, depending on the relative red-shift rates of FH and SH resonances while varying the pump wavelength. Two illustrative experimental examples are provided in Fig.~\ref{fig_3}a and c.
In what we call the `leading case' (Fig.~\ref{fig_3}a), the wavelength at which the maximum absolute value of the frequency offset $\Omega$ occurs at the leading edge of the generated SH power trace, and the SH power continuously decreases as the pump is further tuned into the resonance. We use the vector network analyzer (VNA) technique to probe both the FH and SH detunings (see Methods) \cite{Nitiss2022Opticallyreconfigurablequasi}. The SH resonance is observed to shift away from the generated SH frequency when decreasing the pump frequency (top panel of Fig. \ref{fig_3}a). The underlying dynamic doubly-resonant condition is schematically shown in Fig.~\ref{fig_3}e, where the SH resonance shift rate is regarded as larger than twice the one of the FH resonance ($\mathrm{d}\lambda _s/\mathrm{d}\lambda _{\mathrm{pump}}>2\mathrm{d}\lambda _p/\mathrm{d}\lambda _{\mathrm{pump}}$), considering both thermal and Kerr effects. Initially (stage I), the SH resonance is blue-detuned with respect to the doubled frequency of pump, violating the necessary detuning condition (Eq.~(\ref{deltas})) and thereby no SH generation is possible.
Once the detuning condition is met (stage II), the AOP process is triggered with sufficient pump power, the $\chi^{(2)}$ grating is inscribed and the generated SH power is suddenly increased. Besides, the VNA response is also suddenly triggered. Note that, given the relation between the frequency offset and SH detuning (Eq.~(\ref{Omega})), this minimal detuning condition not only leads to nearly the maximum generated SH power but also the largest frequency shift. The experimental results are in good agreement with our theoretical prediction.
When the pump wavelength is further increased (stages III-IV), the FH detuning reduces slightly but the SH detuning increases significantly (see the VNA response map), explaining the drop of generated SH power accompanied by the decrease in the frequency offset. In the final stage V, the SHG vanishes after the pump gets out of its resonance.

A different behavior is observed in the `trailing case' (Fig.~\ref{fig_3}c). In this case, we observe that the SH resonance approaches the generated SH frequency while increasing the pump wavelength (see the VNA response map). This phenomenology can be attributed to the smaller red-shift rate of the SH mode compared to that of the pump ($\mathrm{d}\lambda _s/\mathrm{d}\lambda _{\mathrm{pump}}<2\mathrm{d}\lambda _p/\mathrm{d}\lambda _{\mathrm{pump}}$). Different from the `leading case', here the required condition Eq.~(\ref{deltas}) is promptly satisfied at the beginning of the pump wavelength tuning. The AOP process is therefore triggered as soon as the threshold pump power (see Supplementary Note I) is reached at stage i, even if the detuning conditions for pump and SH are not optimal.
When the pump is further tuned closer to the resonance (stages ii-iii), the generated SH power grows significantly, owing to the increased proximity to the doubly-resonant condition (Fig.~\ref{fig_3}f). The maximum generated SH power is reached in correspondence with the maximum absolute frequency shift. Further, in the early phase of stage iv, the SH power undergoes a sudden drop as $\delta_s^{\prime}\geqslant 0$, therefore the inscribed nonlinear grating is quickly erased preventing the sustained generation of SH.

To corroborate our interpretation, we replicate our experimental findings through numerical integration of our CME model (see Eqs.~(\ref{CME}) in Methods and Supplementary Note I). For simplicity, we set the effective detunings at pump and SH to change linearly in the simulation, in qualitative agreement with the behavior retrieved from the experimental VNA maps. In the simulation, we intentionally set the pump power to zero when the pump is out of resonance ($\delta_{p}^{\prime} \geqslant 0$), to replicate the experimental triangular-shaped pump transmission \cite{Carmon2004Dynamicalthermalbehavior}. The changes in detunings result in the corresponding variations in the pump transmission, generated SH power and also the frequency offset. Fig.~\ref{fig_3}b and d show the simulation results corresponding to the `leading' and `trailing' cases,  respectively. Numerically we confirm again that the SH can only be generated when the detuning satisfies the condition Eq.~(\ref{deltas}). A small discrepancy between simulation and experiment is the jumps in the pump transmission, typically observed when the \chitwo \ grating is inscribed (erased). Such jumps may be attributed to the change of the pump detuning due to the sudden presence (disappearance) of strong SH inside the microresonator \cite{McKenna2022Ultralowpower}. 

\section*{Discussion and outlook}
\noindent We present in this work the first observation of self-organized spatiotemporal QPM in \SiN \ microresonators. Our experimental findings confirm the existence of a photoinduced traveling \chitwo\ grating at the steady states of the AOP process. The traveling-wave nature manifests itself as a unidirectional sub-kHz Doppler frequency shift to the generated SH, characterized by both self-homodyne and self-heterodyne measurements. For the quasi-static photoinduced electric field, it is worth stressing that, unlike the optical pump and SH fields, it is not associated with a resonant mode as opposed to the hypothesis initially proposed in \cite{Lu2021ConsideringPhotoinducedSecond}, which is contradiction with our experimental observations. First, the resonant mode for the experimentally measured frequency offset, if they existed, would entail a resonance wavelength much larger than the chip itself. In addition, the continuous tuning of the frequency shift $\Omega$ suggests the absence of resonant modes in the sub-kHz range. Instead, our model relies on the first-order CPE dynamics \cite{Dianov1995Photoinducedgenerationsecond,Yakar2022GeneralizedCoherentPhotogalvanic}, where the existence of a characteristic wavelength $\Lambda$ and frequency $\Omega$ emerge respectively from the wavevector mismatch and optimal feedback of the system, leading to a non-stationary steady state.

The phenomenon we observe in this work fits well in the framework of spatiotemporal QPM, bearing similarities with the physics of high-harmonic generation from photoinduced gratings \cite{Zhang2007Quasiphasematching, Cohen2007GratingAssistedPhase}. Our observation bears similarities with the spatiotemporal modulation ubiquitous in inelastic light scattering processes, such as Brillouin and Raman scattering \cite{Gundavarapu2019Subhertzfundamental,Spillane2002UltralowthresholdRaman}, which however entail the interaction between photons and phonons. Analogies can also be drawn with photoinduced space-charge waves by two-wave mixing in photorefractive crystals \cite{Sturman1993Spacechargewaves, Frejlich2007Photorefractivematerialsfundamental}. In this case, moving Bragg gratings can be created by the interference of two frequency-detuned light beams \cite{Refregier1985Two‐beamcouplingphotorefractive} or the application of alternating electric field \cite{Stepanov1985Efficientunstationaryholographic}.
Finally, a similar phenomenon has been observed in nonlinear angular Doppler experiments, where the SHG in rotating nonlinear crystals was shown to follow both energy and angular momentum conservation conditions \cite{Li2016RotationalDopplereffect}.

The developed dynamical model for resonant AOP is of fundamental importance in nonlinear integrated photonics.
It provides a theoretical foundation for  integrated \chitwo\ frequency converters in amorphous materials \cite{Lu2020Efficientphotoinducedsecond, Nitiss2022Opticallyreconfigurablequasi, Clementi2023chipscalesecond,Yakar2023MidIRAll,Li2022Secondharmonicgeneration}, unveiling the necessary detuning conditions for efficient SHG as well as explaining the threshold behavior of the process \cite{Lu2020Efficientphotoinducedsecond,Lu2021ConsideringPhotoinducedSecond,Nitiss2022Opticallyreconfigurablequasi} (see Supplementary Note I). In this perspective, further optimization can be envisaged by leveraging controls over interaction parameters, such as increasing the out-coupling of the SH mode or with a better regulation of the relative mode detuning, now dependent exclusively on pump thermal tuning.
The latter could be improved, for example, with the use of nonlinearly coupled but linearly uncoupled resonators, where the detunings of participating modes could be independently tuned to reach the optimal condition \cite{Menotti2019, Zatti2023Generationphotonpairs, Clementi2024cleo}.
Finally, the developed model can be generalized to \chitwo\ nonlinear processes broadly, such as sum/difference-frequency generation \cite{Yakar2022GeneralizedCoherentPhotogalvanic,Sahin2021}, parametric down-conversion \cite{Dalidet2022perfecttwophoton}, \chitwo-assisted frequency comb generation \cite{Xue2017Secondharmonicassisted,Hu2022Photoinducedcascaded}, comb $f-2f$ self-referencing \cite{Hickstein2019Selforganizednonlinear}, and self-injection locked frequency doubling \cite{Clementi2023chipscalesecond, Li2023Highcoherencehybrid}, among other \chitwo\ functionalities relying on photoinduced nonlinearities.

\vspace{0.3cm}
\noindent\textbf{Methods} \label{method}
\medskip
\begin{footnotesize}

\noindent{\textbf{Theoretical modeling.}} In the absence of pump depletion, the temporal dynamics of AOP-enabled SHG in a microresonator is governed by (see Supplementary Note I):
\begin{equation}
    \begin{aligned}
	\frac{\partial A_p}{\partial t}=&-\left( \frac{\kappa _p}{2}+i\delta _{p}^{\prime} \right) A_p+\sqrt{\kappa _p\eta _pP_{\mathrm{in}}}\\
	\frac{\partial A_s}{\partial t}=&-\left( \frac{\kappa _s}{2}+i\delta _{s}^{\prime} \right) A_s+i\left( \gamma _{spip}A_i \right) A_{p}^{2}\\
	\frac{\partial A_i}{\partial t}=&\beta ^{\prime}\left( A_{p}^{*} \right) ^2A_se^{-i\psi _{\mathrm{ph}}}-\frac{A_i}{\tau}
    \label{CME}
\end{aligned}
\end{equation}
where the subscripts $\left\{ p,s,i \right\}$ are used to denote the intracavity optical pump field, intracavity optical SH field and photoinduced electric field, respectively. Here $|A_{p(s)}|^2=\int_V{d}v\epsilon _{p(s)}\left| E_{p(s)} \right|^2$ and $\left| A_i \right|^2=\frac{1}{2}\int_V{d}v\epsilon _i\left| E_i \right|^2$ represent their modal energy. Note that we include the photoinduced electric field in our coupled mode theory to mathematically describe its temporal dynamics. However, we stress that it is not a resonant mode of the system and it follows instead a first-order CPE dynamics. $\kappa_{p}$ and $ \kappa_{s} $ are the total loss rates (linewidths) of the optical pump and SH fields, $\delta _{p}^{\prime}=\omega _p-\omega_{\text{pump}}$ and $ \delta_{s}^{\prime}=\omega_{s}- 2\omega_{\text{pump}} $ are effective detunings (i.e., considering the Kerr/thermal effects) of the pump ($\omega_{\text{pump}}$) and its SH ($\omega_{\text{SH}}$) from the respective resonance frequencies $\omega _{p}$ and $\omega _{s}$. $\eta_p$ is the pump coupling coefficient, $P_{\mathrm{in}}$ is the pump power in the bus waveguide, $\gamma_{spip}$ is the nonlinear coupling parameter for EFISHG effect, $\beta^{\prime}$ is the effective photogalvanic coefficient incorporating spatial overlap between interacting fields, and $\tau=\epsilon_{i} / \sigma $ is the lifetime of the photoinduced grating (i.e. electric field) with $\epsilon_i$ the static permittivity of the material.

In numerical simulations, the initial condition for initating SHG involves assigning a small value to the photoinduced electric field. For the temporal evolution of fields, the effective detuning $\delta_s^{\prime}$ are set to change linearly with $\delta_p^{\prime}$.
In the steady state, the pump transmission, generated SH power (scaled to the experimental level), and frequency offset are extracted and plotted in Fig.~\ref{fig_3}b and d.

\vspace{0.2cm}
\noindent{\textbf{\SiN \ microresonator.}} 
The \SiN \ microresonator employed in this study is identical to the one with 146 GHz FSR (ring radius of 158 $\mathrm{\mu m}$) used in \cite{Nitiss2022Opticallyreconfigurablequasi}. It is fabricated by LIGENTEC using its AN-technology platform. The bus and ring waveguides have the same cross-section of 1.7 $\times$ 0.5 $\mathrm{\mu m}^2$ and are buried
in SiO$_2$ cladding, supporting a number of transverse electric (TE) modes from $\mathrm{TE_{00}}$ to $\mathrm{TE_{40}}$ in SH band. The loaded Q factors are in the level of  $0.73\times 10^6$ for $\mathrm{TE_{00}}$ resonances at pump wavelengths in C-band.

\vspace{0.2cm}
\noindent{\textbf{Experimental setup.}} 
In the experiment, the light from a tunable continuous-wave laser at telecom C-band is firstly divided into two branches. For the self-heterodyne measurement, the upper branch is frequency shifted by an acousto-optic modulator (AOM) at 92 MHz, then amplified and focused onto a periodically poled KTP crystal for frequency doubling. The generated SH from the crystal, after the pump filtering, serves as a SH frequency reference. 
In the lower branch, the amplified pump is coupled to a \SiN \ microresonator via a lensed fiber for photoinduced SHG. Here the pump wavelength is slowly scanned until the pump and SH become doubly resonant, triggering the AOP process and generating the SH.
The output pump and SH light from the chip are collected using a microscope objective, then separated by a dichroic mirror and measured at respective photodetectors.
The measured SH power is not calibrated thereby lower than its actual generated power level. 
A portion of the generated SH light from the chip and the reference crystal are combined at a beamsplitter, and their optical beating signal is recorded by a ESA at the RBW of 1 Hz. For the self-homodyne measurement, the AOM in the upper branch is bypassed and the temporal beating traces between the SH signals generated from the two branches are recorded by an oscilloscope. 

\vspace{0.2cm}
\noindent{\textbf{Measuring VNA response during the AOP process.}} 
The effective detunings of FH and SH ($ \delta_{p}^{\prime}$ and $\delta_{s}^{\prime}$) are tracked by weakly phase modulating the pump with an electro-optic modulator, and measuring the response with a fast photodetector using a VNA (not shown in Fig. \ref{fig_2}a). The VNA responses are recorded when varying the pump wavelength, and the detailed working principle of the technique can be found in \cite{Nitiss2022Opticallyreconfigurablequasi}. Note that the transmitted pump power ($P_{p}$), generated SH power ($P_{s}$), and the frequency offset ($\Omega/2\pi)$ are measured when the VNA sweeping is off. 

\vspace{0.2cm}
\noindent{\textbf{TPM imaging.}} 
We use the TPM technique to image the \chitwo \ grating inscribed in the Si$_3$N$_4$ microresonator. A Ti:sapphire laser with horizontal polarization is focused on the \chitwo \ grating inscription plane of the poled microreosonator. The focal spot is raster-scanned across the plane and the generated SH signal is collected vertically, thereby obtaining the  \chitwo \ grating images. 

\vspace{0.2cm}
\noindent{\textbf{Data availability}}: The data and code that support the plots within this paper and other findings of this study are available from the corresponding author upon reasonable request.

\vspace{0.2cm}
\noindent{\textbf{Acknowledgements}}:
This work was supported by ERC grant PISSARRO (ERC-2017-CoG 771647) and Swiss National Science Foundation (SNSF grant MINT 214889).

\vspace{0.2cm}
\noindent{\textbf{Author Contributions}}:
J.H. and C.-S.B. conceived the project. J.H., O.Y. and E.N. designed and performed the experiments. J.Z, J.H, M.C. and O.Y. developed the theoretical analysis. The data analysis and numerical simulations were carried out by J.Z., M.C. and J.H. J.Z., M.C., J.H. and C.-S.B. wrote the manuscript with assistance from O.Y. The \SiN \ samples were fabricated by A.S. The project was supervised by C.-S.B.

\vspace{0.3cm}
\noindent{\textbf{Competing interests}}:
The authors declare no competing interests.

\end{footnotesize}

\renewcommand{\bibpreamble}{
$^\ast$These authors contributed equally to this work.\\
$^\dagger${Corresponding author: \textcolor{magenta}{jianqi.hu@epfl.ch}}\\
$^\ddag${Corresponding author: \textcolor{magenta}{camille.bres@epfl.ch}}\\
}

\bibliography{refs.bib}

\end{document}


\newcommand{\SiN}[0]{Si$_3$N$_4$}
\newcommand{\chitwo}[0]{$\chi^{(2)} $ }

\title{Supplementary information for: Self-organized \\ spatiotemporal quasi-phase-matching in microresonators}

\author{Ji Zhou$^{1,\ast}$, 
        Jianqi Hu$^{1,\ast,\dagger}$, 
        Marco Clementi$^{1,\ast}$, 
        Ozan Yakar$^1$, 
        Edgars Nitiss$^{1}$, 
        Anton Stroganov$^{2}$ and Camille-Sophie Brès$^{1,\ddag}$}
\affiliation{
$^1$Photonic Systems Laboratory\char`,{} École Polytechnique Fédérale de Lausanne\char`,{} 1015 Lausanne\char`,{} Switzerland.\\
$^2$LIGENTEC SA\char`,{} EPFL Innovation Park\char`,{} 1024 Ecublens\char`,{} Switzerland.\\
}

\maketitle

\tableofcontents

\setcounter{figure}{0}
\renewcommand{\thefigure}{S\arabic{figure}}
\renewcommand{\theequation}{S\arabic{equation}}

\newpage
\section{Theoretical model}
\noindent In this section, we provide the theoretical framework for describing the temporal dynamics of photoinduced SHG in \SiN\ microresonators. As depicted in Fig. 1c of the main text, this phenomenon arises from the interplay between CPE and EFISHG effect, which are modelled separately in the following.

\subsection{Nonlinear $\chi^{(3)}$ processes in the presence of a DC electric field.}
\noindent The $ \chi^{(3)} $ related nonlinear polarizability $ P_{\text{NL}} $ in the presence of both a static DC field $ E_{\text{dc}} $ and an optical field $ E_{\omega} $ can be expressed as \cite{Friedman2021DemonstrationDCKerr}:
\begin{equation}
    \begin{aligned} P_{\text{NL}} & =\epsilon_{0} \chi^{(3)}\left(E_{\text{dc}}+E e^{-i \omega t}+E^{*} e^{i \omega t}\right)^{3} \\ & =P_{\text{Electrostatic }}+P_{\text {Kerr}}+P_{\text{DC}-\text{Kerr}}+P_{\text{EFISHG }}+P_{\text{THG}}, \end{aligned} 
\end{equation}
with
\begin{equation}
 \left\{\begin{aligned} P_{\text {Electrostatic }} & =\epsilon_{0} \chi^{(3)}\left[E_{\text{dc}}^{3}+6|E|^{2} E_{\text{dc}}\right] \\ P_{\text {Kerr }} & =\epsilon_{0} \chi^{(3)}\left[3|E|^{2} E e^{-i \omega t}+3|E|^{2} E^{*} e^{i \omega t}\right] \\ P_{\text{DC}-\text { Kerr }} & =\epsilon_{0} \chi^{(3)}\left[3 E_{\text{dc}}^{2} E e^{-i \omega t}+3 E_{\text{dc}}^{2} E^{*} e^{i \omega t}\right] \\ P_{\text {EFISHG }} & =\epsilon_{0} \chi^{(3)}\left[3 E_{\text{dc}} E^{2} e^{-i 2 \omega t}+3 E_{\text{dc}}\left(E^{*}\right)^{2} e^{i 2 \omega t}\right] \\ P_{\text{THG}} & =\epsilon_{0} \chi^{(3)}\left[E^{3} e^{-i 3 \omega t}+\left(E^{*}\right)^{3} e^{i 3 \omega t}\right] ,\end{aligned}\right.
\end{equation}
where $\epsilon_0$ is the vacuum permittivity. The polarizability comprises several components, i.e. $P_{\text{Electrostatic}},\\  P_{\text{Kerr}}, P_{\text{DC}-\text{Kerr}}, P_{\text{EFISHG}}$, and $P_{\text{THG}}$, each corresponding to the response of the material at different frequencies. We will foucs our attention on the EFISHG effect, that results in a polarizability oscillating at the SH frequency. Here, in analogy to the regular SHG process, an electric field induced effective $\chi^{(2)}$ can be written as $\chi_{\text{eff}}^{(2)}=3 \chi^{(3)} E_{\text{dc}}$. In the rest of our treatment to the efficient SHG in \SiN \ microresonators, we consider only the relevant Kerr, DC-Kerr, and EFISHG effects.

\subsection{Photoinduced electric field from coherent photogalvanic effect}
\noindent
In this part, the analytical equation for the CPE is presented. 
The first-order CPE involves the generation of an anisotropic coherent current ($ j_{\text{ph}} $) from the multiphoton absorption interference between the fundamental pump and its second harmonic (SH) \cite{Anderson1991Modelsecondharmonic,Dianov1995Photoinducedgenerationsecond,Yakar2022GeneralizedCoherentPhotogalvanic}. In cylindrical coordinates $ (r, \phi, z) $, the generated photogalvanic current $ j_{\text{ph}} $ can be written as:
\begin{equation}
	j_{\text{ph}}(r,\phi ,z)=\beta \left( I_{\text{pump}},I_{\text{SH}} \right) \left[ E_{\text{pump}}^{*}(r,\phi,z) \right] ^2E_{\text{SH}}(r,\phi,z)e^{i\Delta k\phi R-i\psi _{\text{ph}}}+c.c. ,
	\label{j ph}
\end{equation}
where $\beta\left( I_{\text{pump}}, I_{\text{SH}} \right) =\sum_{a,b}{\beta _{ab}}I_{\text{pump}}^{a}I_{\text{SH}}^{b}  (a, b=0,1, \ldots)$ is the generic photogalvanic coefficient with $ I_{\text {pump, SH}}$ the intensities of light, $ E_{\text {pump, SH}} $ are the amplitudes of optical electric fields, $\Delta k= k_{\text{SH}}-2k_{\text{pump}}$ represents the wave vector difference, $ \phi $ denotes the azimuthal angle along the circumference, $ R $ is the resonator radius, $\psi_{\rm{ph}}= \pi$ represents the CPE interaction phase \cite{Yakar2022GeneralizedCoherentPhotogalvanic}. $c.c.$ stands for the complex conjugate.

In the absence of externally applied fields, the photogalvanic current will eventually be compensated by a drift current, which according to Ohm's law takes the form $j_{\text{drift}}=\sigma E_{\text{dc}}$ (note that $j_{\text{ph}}$ and $j_{\text{drift}}$ have opposite directions).
A quasi-static electric field $E_{\text{dc}}$ is therefore established inside the material as a consequence of the CPE, which facilitates the EFISHG process.
When ignoring charge diffusion \cite{Zabelich2022LinearElectrooptic, Yakar2023IntegratedBackwardSecond}, the dynamics of the amplitude of the photoinduced electric field $ E_{\text{dc}} $ can be expressed as \cite{Yakar2022GeneralizedCoherentPhotogalvanic}:
\begin{equation}
 \frac{\partial E_{\text{dc}}(r,\phi ,z)}{\partial t}=\frac{j_{\text{ph}}(r,\phi ,z)}{\epsilon _{\text{dc}}}-\frac{\sigma \left( I_{\text{pump}},I_{\text{SH}} \right)E_{\text{dc}}(r,\phi ,z)}{\epsilon _{\text{dc}}},
	\label{Edc}
\end{equation}
where $ \epsilon_{\text{dc}}=\epsilon_{0} \epsilon_{r} $ represents the permittivity of the medium with $ \epsilon_{r} $ the relative permittivity of the material at low frequency. $\sigma \left( I_{\text{pump}},I_{\text{SH}} \right) =\sum_{a,b}{\sigma _{ab}}I_{\text{pump}}^{a}I_{\text{SH}}^{b}$  is the intensities dependent photoconductivity.

\subsection{Coupled mode equations for resonant SHG.}
\noindent For convenience, in the following we use the subscripts $\left\{ p,s,i \right\}$ to denote the intracavity optical pump field, intracavity optical SH field and photoinduced electric field, respectively.

Based on the combination of CPE and EFISHG effects discussed above, the $\text{SHG} $ in \SiN \space microresonators can be described by the following coupled mode theory \cite{Lin2008proposalhighlytunable, Lu2021ConsideringPhotoinducedSecond}:
\begin{equation}
	\left\{ \begin{aligned}
		\frac{\partial A_p(t)}{\partial t}= & -\left( \frac{\kappa _p}{2}+i\delta _p \right) A_p+\sqrt{\kappa _p\eta _pP_{\text{in}}}                                                                                             \\
		                                    & +i\left( \gamma _{pppp}\left| A_p \right|^2+2\gamma _{psps}\left| A_s \right|^2+2\gamma _{pipi}\left| A_i \right|^2 \right) A_p+i\left( 2\gamma _{pspi}A_{i}^{*} \right) A_sA_{p}^{*} \\
		\frac{\partial A_s(t)}{\partial t}= & -\left( \frac{\kappa _s}{2}+i\delta _s \right) A_s                                                                                                                                    \\
		                                    & +i\left( \gamma _{ssss}\left| A_s \right|^2+2\gamma _{spsp}\left| A_p \right|^2+2\gamma _{sisi}\left| A_i \right|^2 \right) A_s+i\left( \gamma _{spip}A_i \right) A_{p}^{2}           \\
		\frac{\partial A_i(t)}{\partial t}= & \beta ^{\prime}\left( A_{p}^{*} \right) ^2A_se^{-i\psi _{\text{ph}}}-\frac{A_i}{\tau},                                                                                               \\
	\end{aligned} \right.
	\label{CME}
\end{equation}
where $|A_{p,s}|^2=\int_V{d}v\epsilon _{p,s}\left| E_{p,s} \right|^2$ represent the modal energy of optical pump and SH fields, $\left| A_i \right|^2=\frac{1}{2}\int_V{d}v\epsilon _i\left| E_i\right|^2$ is also the energy of the photoinduced electric field considering its spatial distribution despite not being a resonant mode (see Eq.~(\ref{j ph}, \ref{Edc})). $\epsilon _{p,s}=\epsilon_0n_{p,s}^2$ with $n_{p(s)}$ the refractive indices of \SiN\ waveguide for pump and $\text{SH}$. 

For the other parameters in Eq.~(\ref{CME}), $ \kappa_{p} $ and $ \kappa_{s} $ are the total loss rates of optical pump and $ \text{SH} $ fields, respectively. $ \delta_{p}=\omega_{p}-\omega_{\text {pump}}$ and $ \delta_{s}=\omega_{s}-2\omega_{\text {pump}} $ are the detunings for pump resonance $ \left(\omega_{p}\right) $ and the $ \text{SH} $ resonance $ \left(\omega_{s}\right)$. $\eta_{p}=\kappa_{\text{ex}, p}/\kappa_{p} $ represents the coupling efficiency of the pump field with $ \kappa_{\text{ex}, p} $ the output coupling rate. $ \gamma_{j k l m}\left(E_{j}, E_{k}, E_{l}, E_{m}=E_{p}, E_{s}, E_{i}^{*}\right) $ denote the third-order nonlinear parameters defined in a similar manner as in \cite{Lin2008proposalhighlytunable,Lu2021ConsideringPhotoinducedSecond}. $\tau=\epsilon_{i} / \sigma $ is the lifetime of photoinduced electric field or inscribed nonlinear grating. Additionally, $ \beta^{\prime}=\frac{\xi}{\epsilon_{p} \sqrt{2 \epsilon_{s} \epsilon_{i}}} \beta $ with $\xi$ the nonlinear overlap defined as: 
\begin{equation}
\xi =\sqrt{\frac{\int_V{d}v\left| \left( E_{p\,\,}^{*} \right) ^2E_s \right|^2}{\left( \int_V{d}v\left|  E_{p\,\,}  \right|^2 \right) ^2\int_V{d}v\left| E_s \right|^2}}.
\end{equation}

In Eq. (\ref{CME}), the nonlinear terms in the first two equations describe self-phase modulation (SPM), cross-phase modulation (XPM), DC-Kerr and EFISHG effects, respectively. Derived from Eq. (\ref{Edc}), the third equation describes the dynamics of the photoinduced electric field. It is worth noting that in this context, QPM is inherently satisfied among these three fields, the phase mismatch terms can be reduced by variable substitution.

The \SiN\ microresonantor utilized in this work is identical to the one (with FSR of 146 GHz) in \cite{Nitiss2022Opticallyreconfigurablequasi}, and we use the same material parameters from \cite{Yakar2022GeneralizedCoherentPhotogalvanic,Nitiss2020FormationRulesDynamics}. Table \ref{sigma beta table} lists the basic information of the \SiN\ microresonator as well as the intensity-dependent photogalvanic coefficients and photoconductivity coefficients used in the simulation.
\begin{table}[h]
	\centering
	\caption{\centering Parameters used in simulations.}
	\label{sigma beta table}
	\begin{threeparttable}
		\begin{tabular}{cc}
			\hline
			Parameters                                                                       & Value                                                                                                                                                                                                                                                                                            \\ \hline
			microresonantor cross-section, $h \times w $                                  & $1.7\times 0.5~\mu \rm{m^2}$                                                    \\                                         ring radius, $R$     &             $158~\mu \rm{m}$                                                                                             \\
        third-order susceptibility \cite{Lu2021ConsideringPhotoinducedSecond}, $\chi^{(3)}_{spip}$                                              & $3.39\times 10^{-21}~\mathrm{m^2/V^2}$                                                                                                                                                                                                             \\

			 CPE interaction phase, $\psi_{\rm{ph}}$                                                & $\pi$ rad                                                                                                                                                                                                                                                                                            \\
			photoconductivity coefficients \cite{Yakar2022GeneralizedCoherentPhotogalvanic}, $\sigma_{mn} $                                  & $\begin{cases}  \sigma_{00}\approx 5.12 \times 10^{-17}\ \mathrm{S}\cdot \text{m}^{-1} \ \tnote{*} \\\sigma_{02}=1.19\times 10^{-33} \ \mathrm{S \cdot m^3 \cdot W^{-2}}  \\ \sigma_{21}=1.36\times 10^{-48} \ \mathrm{S\cdot m^5 \cdot W^{-3}}
				                                                                                    \end{cases}$ \\
			photogalvanic coefficients \cite{Yakar2022GeneralizedCoherentPhotogalvanic}, $\beta_{mn}$                                         & $\begin{cases} \beta_{01}= 8.22 \times 10^{-38} \ \rm{m^3 \cdot V^{-4} }     \\ \beta_{20}= 6.17 \times 10^{-53} \ \rm{ m^5 \cdot V^{-4} \cdot W^{-1}}   \end{cases}$
			\\
		\hline
		\end{tabular}
		\begin{tablenotes}
			\item[*]$\sigma _{00}=\epsilon _{i}/\tau_{00} $ with $\tau_{00}$ being approximately 15 days \cite{Nitiss2020FormationRulesDynamics}.
		\end{tablenotes}
	\end{threeparttable}
\end{table}

\subsection{Traveling grating and AOP condition}
\noindent An intriguing property of the system descried by Eq. (\ref{CME}) is that it can sustain a nonstationary steady state, which implies that the photoinduced electric field is traveling at the equilibrium. We can analyze the system by writing the Jacobian matrix of Eq. (\ref{CME}) as:
\begin{equation}
 \boldsymbol{M}=\left(\begin{array}{cccccc}-\left(\frac{\kappa_{p}}{2}+i \delta_{p}^{\prime}\right) & 0 & 0 & 0 & 0 & 0 \\ i 2 \gamma_{spip} A_{p} A_{i} & -\left(\frac{\kappa_{s}}{2}+i \delta_{s}^{\prime}\right) & i \gamma_{s p i p} A_{p}^{2} & 0 & 0 & 0 \\ 0 & \beta^{\prime}(A_{p}^{*})^{2} e^{-i \psi_{\text {ph }}} & -\frac{1}{\tau} & 2 \beta^{\prime} A_{p}^{*} A_{s} e^{-i \psi_{\text {ph }}} & 0 & 0 \\ 0 & 0 & 0 & -\left(\frac{\kappa_{p}}{2}-i \delta_{p}^{\prime}\right) & 0 & 0 \\ 0 & 0 & 0 & -i 2 \gamma_{s p i p} A_{p}^{*} A_{i}^{*} & -\left(\frac{\kappa_{s}}{2}-i \delta_{s}^{\prime}\right) & -i \gamma_{s p i p}\left(A_{p}^{*}\right)^{2} \\ 2 \beta^{\prime} A_{p} A_{s}^{*} e^{i \psi_{\text {ph }}} & 0 & 0 & 0 & \beta^{\prime} A_{p}^{2} e^{i \psi_{\text{ph}}} & -\frac{1}{\tau}\end{array}\right). 
\end{equation}
Here we group all the nonlinear phase terms in Eq.~(\ref{CME}) to the effective detunings $\delta_p^{\prime},\delta_s^{\prime}$), which can also include the thermo-optic effects, though not specifically described in Eq. (\ref{CME}).

The characteristic equation for the eigenvalues of $\boldsymbol{M}$ is:
\begin{equation}
 \begin{array}{l}{\left[\left(\frac{\kappa_{p}}{2}+i \delta_{p}^{\prime}\right)+\lambda\right]\left\{\left[\left(\frac{\kappa_{s}}{2}+i \delta_{s}^{\prime}\right)+\lambda\right]\left(\frac{1}{\tau}+\lambda\right)+i \varUpsilon\left|A_{p}\right|^{4}\right\}} \\ \cdot\left[\left(\frac{\kappa_{p}}{2}-i \delta_{p}^{\prime}\right)+\lambda\right]\left\{\left[\left(\frac{\kappa_{s}}{2}-i \delta_{s}^{\prime}\right)+\lambda\right]\left(\frac{1}{\tau}+\lambda\right)-i \varUpsilon\left|A_{p}\right|^{4}\right\}=0, \end{array} 
\label{eigen}
\end{equation}
where $\varUpsilon  =\gamma _{spip}\beta ^{\prime}$.

Under the assumptions of $\left| i\varUpsilon A_p  ^4 \right|\ll \left| \frac{\kappa _s}{2}+i\delta _{s}^{\prime} \right|$ and $\frac{1}{\tau}\ll \left| \frac{\kappa _s}{2}+i\delta _{s}^{\prime} \right|$ (the photoinduced electric field exhibits much slower growth or decay rates compared to the loss rates and detunings of optical fields), the first three eigenvalues from the first two terms of Eq. (\ref{eigen}) are given by:
\begin{equation}
 \left\{\begin{array}{l}\lambda_{1}=-\left(\frac{\kappa_{p}}{2}+i \delta_{p}^{\prime}\right) \\ \lambda_{2} \approx-\left(\frac{\kappa_{s}}{2}+i \delta_{s}^{\prime}\right) \\ \lambda_{3} \approx-\frac{1}{\tau}-\frac{i \varUpsilon\left|A_{p}\right|^{4}}{\kappa_{s} / 2+i \delta_{s}^{\prime}}=-\frac{\varUpsilon\left|A_{p}\right|^{4} \delta_{s}^{\prime}}{\kappa_{s}^{2} / 4+\left(\delta_{s}^{\prime}\right)^{2}}-\frac{1}{\tau}-i \frac{\varUpsilon\left|A_{p}\right|^{4} \kappa_{s} / 2}{\kappa_{s}^{2} / 4+\left(\delta_{s}^{\prime}\right)^{2}},\end{array}\right. 
 \label{lambda123}
\end{equation}

The nonvanishing imaginary part of $\lambda_3$ gives the oscillation frequency ($\Omega$) of the photoinduced field:
\begin{equation}
		\text{Im}\left\{ \lambda _3 \right\} \approx -\frac{\varUpsilon \left| A_p \right|^4\kappa _s/2}{\kappa _{s}^{2}/4+\left( \delta _{s}^{\prime} \right) ^2}=\Omega<0.
	\label{Omega}
\end{equation}

The additional relation between $ \tau $ and $ \Omega $ is given by the periodic solutions at steady state. The existence of non-stationary steady states is not theoretically proved here but  consistently observed experimentally, we hence assume the form of the solution of Eq. (\ref{CME}) at equilibrium is $ A_{p} \rightarrow A_{p}, A_{s} \rightarrow A_{s} e^{i \Omega t}, A_{i} \rightarrow A_{i} e^{i \Omega t}$, from Eq.~(\ref{CME}) we obtain:
\begin{equation}
    \left\{ \begin{array}{l}
	A_p=\frac{\sqrt{\kappa _p\eta _pP_{\text{in}}}}{\left( \frac{\kappa _p}{2}+i\delta _{p}^{\prime} \right)}\\
	i\Omega A_s=-\left( \frac{\kappa _s}{2}+i\delta _{s}^{\prime} \right) A_s+i\left( \gamma _{spip}A_i \right) A_{p}^{2}\\
	i\Omega A_i=\beta ^{\prime}\left( A_{p}^{*} \right) ^2A_se^{-i\psi _{\text{ph}}}-\frac{A_i}{\tau}.
\end{array} \right. 
\label{ansatz}
\end{equation}
This leads to
\begin{equation}
\left\{ \begin{aligned}
	\ & \frac{\kappa _s}{2\tau}-\left( \delta _{s}^{\prime}+\Omega \right) \Omega =0\\
	&\frac{\delta _{s}^{\prime}+\Omega}{\tau}+\Omega \frac{\kappa _s}{2}=-\Upsilon \left| A_p \right|^4,
\end{aligned} \right.
\end{equation}
In the limit where $\left| \delta _{s}^{\prime} \right|\gg \left| \Omega \right|$, we can derive
\begin{equation}
	\frac{1}{\tau} \approx -\frac{\varUpsilon\left|  A_p \right|^4 \delta_{s}^{\prime}}{\kappa_{s}^{2} / 4+\left(\delta_{s}^{\prime}\right)^{2}}=\frac{2 \delta_{s}^{\prime}}{\kappa_{s}} \Omega. 
	\label{tau}
\end{equation}

The relation $\Omega \approx \frac{\kappa_{s}}{2 \tau \delta_{s}^{\prime}}$ in Eq.~(\ref{tau}) is used to fit the simulation results, as illustrated in Fig.~3 in the main text, and Fig.~\ref{zero crossing}(b) as well as Fig.~\ref{other exps}(a). More importantly, we can see that the condition for efficient SHG is $ \delta_{s}^{\prime}<0 $ as $ \tau $ and $\kappa_s$ are positive, while $\Omega$ remains negative.

In addition, from Eq.~(\ref{lambda123}) and Eq.~(\ref{tau}), we notice that $\text{Re}\{\lambda _3\}\approx 0$ holds for the ansatz used in Eq.~(\ref{ansatz}), explaining the constant-power steady states reported in this work. Experimentally, we also observe other non-stationary states with periodic oscillations in SH power, likely due to the intricate CPE or the thermal effect. We leave the detailed study for future work.

\subsection{Pump threshold and conversion efficiency}
\noindent Different from conventional threshold-less cavity-enhanced SHG \cite{Guo2016Secondharmonicgeneration}, the photoinduced SHG typically exhibits a threshold \cite{Lu2020Efficientphotoinducedsecond,Lu2021ConsideringPhotoinducedSecond,Nitiss2022Opticallyreconfigurablequasi}. This threshold behavior can be attributed, in part, to the requirement for the inscribed $ \chi^{(2)} $ grating from CPE to surpass its erasure mediated by conductivity. Here we simulate the SHG threshold using the dark conductivity $\sigma_{00}$ of the material, while omitting for simplicity the potential contributions of the photoconductivity from the pump (e.g., $\sigma_{40}$). 

When pumped above the threshold power, the system loses its stability and evolves from the state without SHG $ \left(A_{s}=A_{i}=0\right) $ to the state that generates SH. Mathematically, it writes:
\begin{equation}
	\text{Re}\left\{ \lambda _3 \right\} \approx -\frac{\varUpsilon \left| A_p  \right|^4\delta _{s}^{\prime}}{\kappa _{s}^{2}/4+\left( \delta _{s}^{\prime} \right) ^2}-\frac{1}{\tau}\geqslant 0.
	\label{lambda3}
\end{equation}

As the parameters $\varUpsilon  =\gamma _{spip}\beta ^{\prime}$ and $ \tau= \epsilon_{i} / \sigma $ depend on the light intensities (see Table \ref{sigma beta table}), with the relation $A_p=\frac{\sqrt{\kappa _p\eta _pP_{\text{in}}}}{\kappa _p/2+i\delta _{p}^{\prime}}$ the pump threshold power in the bus waveguide can be derived as: 
\begin{equation}
	P_{\text{th}}(\delta _{p}^{\prime},\delta _{s}^{\prime})=\sqrt[4]{\frac{\sigma _{00}\left( \kappa _{p}^{2}/4+\left( \delta _{p}^{\prime} \right) ^2 \right) ^4\left( \kappa _{s}^{2}/4+\left( \delta _{s}^{\prime} \right) ^2 \right)}{\mathcal{L}\delta _{s}^{\prime}}},
	\label{Pth}
\end{equation}
where $\mathcal{L} =\frac{\gamma _{spip}\xi \kappa _{\text{ex},p}^{4}\beta _{20}e^{-i\psi _{\text{ph}}}}{\epsilon _p}\sqrt{\frac{\epsilon _i}{2\epsilon _s}}\left( \frac{D_{1,p}}{2\pi A_{\text{eff} ,p}} \right) ^2$ with $D_{1,p}/2\pi$ and $ A_{\text{eff}, p} $ representing the FSR and effective mode area of the pump field, respectively.
Note that the pump threshold is minimized for $(\delta _{p}^{\prime},\delta _{s}^{\prime})=(0,-\frac{\kappa _s}{2})$, which represents the optimal operating point.

\begin{figure*}[htbp]
	\centering
	\includegraphics[width=1\linewidth]{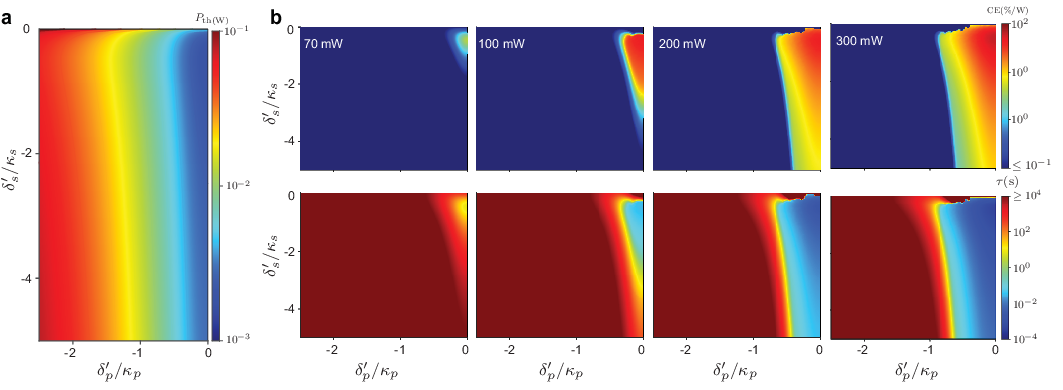}
	\caption{ \textbf{Simulation of threshold behavior of photo-induced SHG.}
 \textbf{a}, Simulated threshold power $P_{\text{th}}$ as a function of pump and SH detunings. 
			  \textbf{b}, Conversion efficiency (CE) map (top) and grating lifetime $ \tau $ (bottom) as  functions of detunings under different pump power. 
			  In simulation, the $\text{TE}_{00}$ FH mode (around $ 1542.7 \text{~nm} $) and  $ \text{TE}_{30}$ SH mode and are considered \cite{Nitiss2022Opticallyreconfigurablequasi}, with simulation parameters $Q_{p}\left(Q_{s}\right)=0.73~(1.30) \times 10^{6}, \kappa_{p}\left(\kappa_{s}\right) / 2 \pi=270~(300)~ \text{MHz}, \eta_{p}\left(\eta_{s}\right)=0.5~(0.5)$ used.
	}\label{maps}
\end{figure*}

The simulated threshold power as a function of detunings is graphically illustrated in Fig.~\ref{maps}(a). The threshold power exhibits a stronger dependence on pump detuning $ \delta_{p}^{\prime} $ than $ \text{SH} $ detuning $ \delta_{s}^{\prime} $, and it could vary approximately an order of magnitude when the pump detuning is changed over a linewidth at pump. Based on the simulation, we estimate that the detuning-dependent threshold power ranges from $1 \times 10^{-3}$ to $2 \times 10^{-2}$ W, which roughly aligns with the previously observed value of approximately $2.5 \times 10^{-2}$ W \cite{Nitiss2022Opticallyreconfigurablequasi}. Table \ref{threshold power table} summarizes the comparison of threshold power observed in different studies. In microresonators, the threshold pump power $P_{\text{th}}$ scales inversely with the finesse $\mathcal{F}=D_{1,p}/\kappa_p$ assuming optimal detuning conditions, as derived from Eq.~(\ref{Pth}).

The upper panel in Fig.~\ref{maps}(b) shows the simulated conversion efficiency $ \text{CE}=P_{s,\text{out}} / P_{p,\text{in}}^{2} $ with $ P_{s,\text {out}} $ and $ P_{p, \text { in }} $ being the generated $ \text{SH} $ and the pump power in the bus waveguide, respectively. The lifetime $ \tau $ as a function of detunings is also plotted in the bottom panel. Similar to the saturation behavior of CE in conventional SHG \cite{Guo2016Secondharmonicgeneration}, the CE in our system is clamped at approximately $ 60 \% / \text{W} $, given the simulation parameters listed in the caption. Note that in our simulations, the back conversion from SH to pump leading to pump depletion is ignored. The saturation behavior of CE is attributed to the coherent photogalvanic effect. 

Qualitatively, as the intracavity optical power increases, both the photogalvanic coefficient $\beta$ (related to the `gain' of the photoinduced electric field $A_{i}$) and conductivity $\sigma $ (`loss' of $A_{i}$) will increase. Above a certain level, further increase in the pump power will contribute more to the photoconductivity than to the photogalvanic coefficient and eventually clamp the CE.
\begin{table}[htbp]
	\centering
	\caption{\centering Comparison of the threshold power for photoinduced SHG in different works.}
	\label{threshold power table}
	\begin{threeparttable}
		\begin{tabular}{ccc}
			\hline
			References                                           & Pump finesse, $ \mathcal{F} $ & $ P_{\text {th }}(\text{W})$        \\ \hline
   microring \cite{Nitiss2022Opticallyreconfigurablequasi} & $ 545 $                            & $< 2.5 \times 10^{-2} $       \\
			microring \cite{Lu2020Efficientphotoinducedsecond}      & $ 6472 $                           & $ <4 \times 10^{-3} $                 \\
			microring \cite{Lu2021ConsideringPhotoinducedSecond}    & $ 5000 $                           & $ 2.3 \times 10^{-3} $                \\
			single-pass waveguide \cite{Porcel2017Photoinducedsecond}      & $ - $                              & $ <9.84$                               \\
			single-pass waveguide \cite{Nitiss2020FormationRulesDynamics}  & $ - $                              & $ <38 $                               \\
			single-pass waveguide \cite{Billat2017Largesecondharmonic}     & $ - $                              & $ <60 $                               \\
			\hline
		\end{tabular}
		\begin{tablenotes}
			\item[*]  Additional comparisons are available in \cite{Lu2021ConsideringPhotoinducedSecond}.
		\end{tablenotes}
	\end{threeparttable}
\end{table}

\section{SHG perturbed by a neighboring SH mode}
\label{zerocrossing}

\begin{figure*}[h!]
	\centering
	\includegraphics[width=1\linewidth]{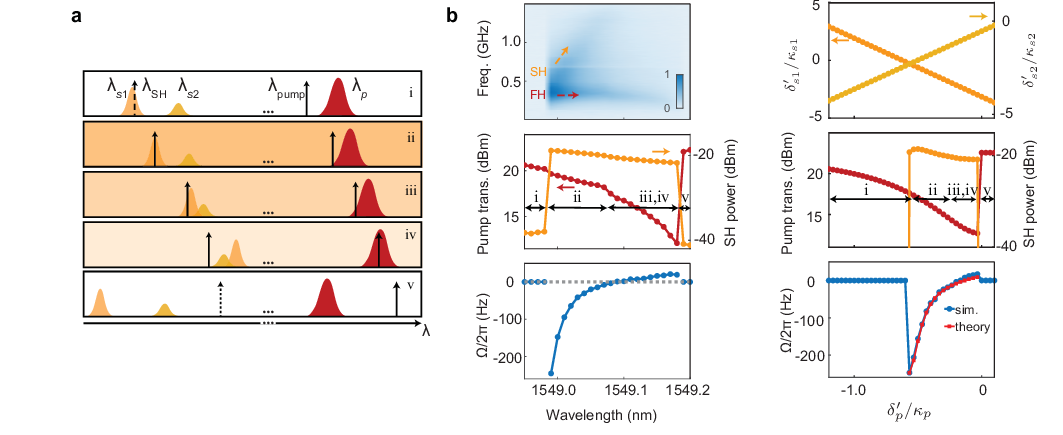}
		\caption{\textbf{Dynamics of AOP-enabled SHG in the presence of a neighboring SH mode.} 
    \textbf{a}, Schematic illustration of the pump and SH resonance distributions with respect to the pump and SH signals, respectively. 
    \textbf{b}, Experimental (left) and simulation (right) results. $ \delta_{s1}^{\prime} $ and $ \delta_{s2}^{\prime} $ denote the effective detunings of the main and the perturbing SH modes, respectively. In the simulation, the frequency offset is also fitted by the theoretical prediction $\Omega = \text{Im}\left\{\lambda_4\right\}$. 
	}\label{zero crossing}
\end{figure*}

\noindent Experimentally, there are instances where the frequency offset is larger than zero as shown in the Fig.~2b in the main text, which is not supported by the theory we establish above (Eq. (\ref{Omega})). A possible explanation is the participation of another neighboring SH mode in the SHG process. Considering two following facts: i) from the VNA response maps in Fig.~3a and c of the main text and also Fig.~\ref{zero crossing}(b), it is evident that during SHG, the pump is thermally locked to its nearest resonance (see the nearly horizontal FH branch), while the generated SH is not (the tilted SH branch) \cite{Nitiss2022Opticallyreconfigurablequasi}. This implies that the SH detuning could shift a lot by several linewidths. ii) In a multimode \SiN\ waveguide, a multitude of transverse modes at SH band show distinct thermo-optic coefficients, and it is possible that sometimes a neighboring SH mode becomes more resonant during the frequency tuning at pump. These reasonable assumptions can lead to simultaneous SHG for multiple SH modes, which may explain the phenomenon of $\Omega>0$ we observe in the experiment.

Here we consider the scenario where the electric field is composed of two components ($A_{i,q}$). In this case, Eq. (\ref{CME}) can be extended as:
\begin{equation}
	\left\{ \begin{aligned}
		\frac{\partial A_p}{\partial t}= & -\left( \frac{\kappa _p}{2}+i\delta _{p}^{\prime} \right) A_p+\sqrt{\kappa _p\eta _pP_{\text{in}}}               \\
		\frac{\partial A_s}{\partial t}= & -\left( \frac{\kappa _s}{2}+i\delta _{s}^{\prime} \right) A_s+i\left( \gamma _{spip}A_i \right) A_{p}^{2}          \\
		\frac{\partial A_i}{\partial t}= & \beta _{s}^{\prime}\left( A_{p}^{*} \right) ^2A_se^{-i\psi _{\text{ph}}}-\frac{A_i}{\tau _i}+ig^*A_q             \\
		\frac{\partial A_w}{\partial t}= & -\left( \frac{\kappa _w}{2}+i\delta _{w}^{\prime} \right) A_w+i\left( \gamma _{wpqp}A_q \right) A_{p}^{2}e^{i\psi} \\
		\frac{\partial A_q}{\partial t}= & \beta _{w}^{\prime}\left( A_{p}^{*} \right) ^2A_we^{-i\psi _{\text{ph}}}-\frac{A_q}{\tau _q}+igA_i,               
	\end{aligned} \right.
	\label{5eqs}
\end{equation}
where $\{A_p, A_s, A_i\}$ and $\{A_p, A_w, A_q\}$ are QPM mode pairs, $g$ is the linear coupling coefficient related to the charge diffusion, $\psi$ is the spatial phase difference between the two electric field components.

For simplicity, we ignore the linear coupling between photoinduced electric fields ($g=0$). Generally, one SH mode ($A_s$) dominates the SHG process and the other one ($A_w$) can be considered as a perturbation. To simplify the simulation, we assume the ratio between the two electric field strengths is fixed ($\alpha = A_q/A_i \ll 1 $), such that Eq. (\ref{5eqs}) can be reduced to four equations by adding the two photoinduced electric field equations:
\begin{equation}
	\left\{ \begin{aligned}
		\frac{\partial A_p}{\partial t}= & -\left( \frac{\kappa _p}{2}+i\delta _{p}^{\prime} \right) A_p+\sqrt{\kappa _p\eta _pP_{\text{in}}}                                                                   \\
		\frac{\partial A_s}{\partial t}= & -\left( \frac{\kappa _s}{2}+i\delta _{s}^{\prime} \right) A_s+i\left( \gamma _{spip}A_i \right) A_{p}^{2}                                                              \\
		\frac{\partial A_w}{\partial t}= & -\left( \frac{\kappa _w}{2}+i\delta _{w}^{\prime} \right) A_w+i\left[ \gamma _{wpqp}\left( \alpha A_i \right) \right] A_{p}^{2}e^{i\psi}                                             \\
		\frac{\partial A_i}{\partial t} \approx & \beta _{s}^{\prime}\left( A_{p}^{*} \right) ^2A_se^{-i\psi _{\text{ph}}}+\beta _{w}^{\prime}\left( A_{p}^{*} \right) ^2A_we^{-i\psi _{\text{ph}}}-\frac{A_i}{\tau}, \\
	\end{aligned}\right.
	\label{mode interaction 4eqs}
\end{equation}
where $1/\tau =  1/\tau _i+\alpha/\tau _q$ and 
we set $\alpha=0.1$ in our simulation.

Similar to the derivation of Eq.~(\ref{lambda123}), we can obtain the expression of the eigenvalue related to the evolution of photoinduced electric field at the steady state:
\begin{equation}
	\lambda_4 \approx -\frac{1}{\tau}-\frac{i\alpha\varUpsilon _w\left| A_p \right|^4e^{i\psi}}{\kappa _{w}/2+i\delta _{w}^{\prime}}-\frac{i\varUpsilon _s\left| A_p \right|^4}{\kappa _s/2+i\delta _{s}^{\prime}}.
 \label{lambdash12}
  \end{equation}
We notice that the eigenvalue in Eq.~(\ref{lambdash12}) is now perturbed by the adjacent SH mode, as compared to Eq.~(\ref{lambda123}). In the simulation shown in Fig.~\ref{zero crossing}, we assume $\psi$ takes the value of $\pi/2$ and calculate the frequency offset by $\Omega=\text{Im}\left\{\lambda_4\right\}$ for the fitting.

Figure~\ref{zero crossing} shows the simulated zero-crossing behavior of the frequency offset. The presence of two neighboring different order modes at $ \text{SH} $ band is considered. When tuning the pump wavelength, the $ \text{SH} $ mode $ \lambda_{\text{s} 2} $ will start to effectively perturb the existing SH generated from the mode $ \lambda_{\text{s} 1} $ at a certain pump wavelength, resulting in $\Omega >0$. In addition, such a zero-crossing behavior also exhibits pump power dependence (see Supplementary Note III for more details).

Further, we  anticipate that if the detuning of the neighboring SH mode becomes small enough, competition between the two modes may occur, until reaching a stable state where the adjacent mode takes over the SHG. Such a phenomenon has already been observed previously (Fig.~S8 in \cite{Nitiss2022Opticallyreconfigurablequasi}) and further investigation is left for future work.

Apart from the perturbation from a neighboring SH mode considered above, sum-frequency generation \cite{Hu2022Photoinducedcascaded}, third-harmonic generation \cite{Levy2011Harmonicgenerationsilicon} and other nonlinear processes may also lead to the zero-crossing behavior through nonlinear mode interaction.

\section{Power dependence} \label{Power dependence}
\begin{figure}[h!]
	\centering
	\includegraphics[width=0.5\linewidth]{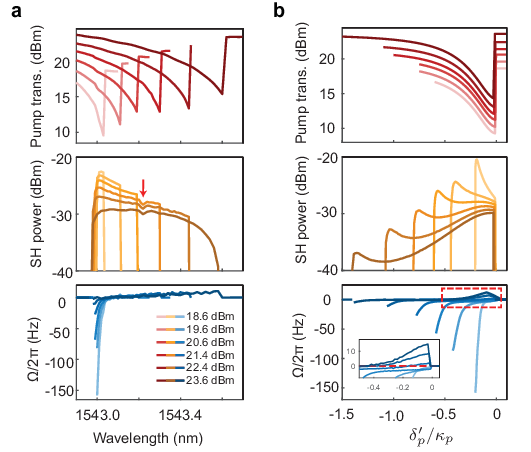}
	\caption{\textbf{SHG characteristics at the resonance near 1543.0 nm under different pump power}. \textbf{a}, Top: Measured pump transmission; Middle: Measured generated SH power; Bottom: Measured frequency offsets. In the experiments, the generated SH power decreases when increasing the pump power or tuning the pump into the resonance. Above a certain pump power level, discrete changes in the generated SH power (denoted by the red arrow) and the zero-crossing of the frequency offset are observed. \textbf{b}, Top: Simulated pump transmission; Middle: Simulated generated SH power; Bottom: Simulated frequency offsets. In the simulation, the mode perturbation is introduced to the cases with pump power levels of 21.4, 22.4, and 23.6 dBm to reproduce the zero-crossing behaviors, as shown in the inset. 
		}\label{power dependence}
\end{figure}

\noindent In this section, we show the power-dependent behaviours of the photo-induced SHG. Figure~\ref{power dependence} shows the measurement results under different pump power levels for the resonance near 1543.0 nm, which belongs to the leading case. It can be seen that higher pump power here corresponds to less efficient SHG, as shown in the left upper and middle panels in Fig.~\ref{power dependence}(a). Note that at the same pump wavelength, the generated SH power decreases with increasing pump power. This is because higher pump power shifts the SH resonance farther away from the generated SH frequency via the thermal and Kerr effects. In addition, given that the photoconductivity depends more on the intensity of SH than that of pump \cite{Yakar2022GeneralizedCoherentPhotogalvanic}, the grating lifetime would decrease with the generated SH power, i.e., $\tau \propto \left( \sigma _{00}+\sigma _{02}I_{\text{SH}}^{2}+\sigma _{21}I_{\text{pump}}^{2}I_{\text{SH}} \right)^{-1}$. The combined effects explain the experimental observations that lower pump power corresponds to a higher frequency shift since $\Omega \propto \frac{1}{\delta _{s}^{\prime}\tau}$. 
 
Experimentally, the zero crossing behaviour of the measured frequency offset becomes increasingly evident at high pump power. As discussed in Supplementary Note II, we attribute the zero crossing to the perturbation of a neighboring SH mode, which are present in Fig.~2b of the main text as well. The small drop in the generated SH power trace, marked by the red arrow in Fig.~\ref{power dependence}(a), may be a signature of the change of the rotating direction of the grating when the frequency offset $\Omega$ changes the sign. In Fig.~\ref{power dependence}(b), we perform simulations to reproduce the experimental results. In the simulation, perturbations from a neighboring lossy resonance with a large linewidth are taken into account for the cases using 21.4, 22.4, and 23.6 dBm pump power. It can be seen that the frequency offset zero-crossing behaviors up to few Hz are qualitatively reproduced in the simulation.

\section{Other experimental results}

\begin{figure*}[h]
	\centering
	\includegraphics[width=1\linewidth]{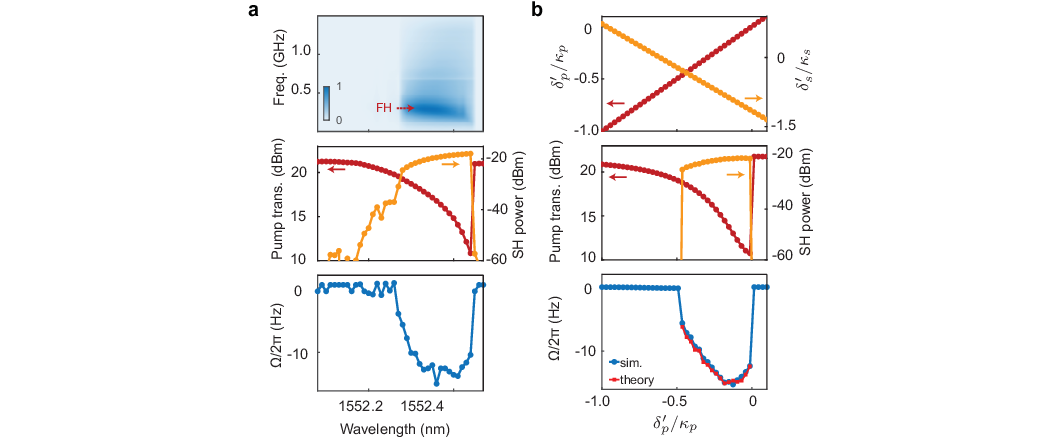}
	\caption{\textbf{Additional measurement and simulation results of photoindcued SHG in the \SiN \space microresonator}. For the AOP around $1552.2$ nm, the maximum absolute value of the frequency offset ($\Omega_{\max}$) occurs at the middle of the SH power trace. \textbf{a}, Top: VNA response map; Middle: Measured pump transmission and generated SH power; Bottom: Measured frequency offset ($ \Omega / 2 \pi $). The experimental non-sharp leading edge of SH power trace may be attributed to SHG from a preceding SH mode or less efficient SHG from a not fully erased \chitwo\ grating. 
    \textbf{b}, Top: the FH and SH effective detunings setting in the simulation; Middle: Simulated pump transmission and  generated SH power; Bottom: Simulated frequency offset and the theoretical prediction based on Eq. (\ref{tau}). 
	}\label{other exps}
\end{figure*}

\noindent Apart from the experimental results that correspond to the leading and trailing cases in the main text, there are also other scenarios with one of the examples shown in Fig.~\ref{other exps}. Fig.~\ref{other exps} illustrates the AOP dynamics for the pump wavelength around $1552.2$ nm, where the position of the maximum absolute value of the frequency offset does not align with either the leading or the trailing edge of the generated SH power trace. This behavior, however, does not contradict the presented theory (Eq. (\ref{tau})), as the frequency offset does not strictly follow an inversely proportional relation to the SH detuning. This is because the grating lifetime $\tau$, which depends on power, is also related to the detuning. In this specific case, for which the participation SH mode is identified by TPM imaging as TE$_{40}$, we expect that $\tau$ becomes significant given that this resonance corresponds to the highest SH power \cite{Nitiss2022Opticallyreconfigurablequasi} and hence start significantly influence the frequency offset.

\bibliography{refs.bib}